\documentclass[conference,a4paper,final,twocolumn,10pt,twoside]{IEEEtran}
\usepackage{layout}
\IEEEoverridecommandlockouts 

\ifCLASSINFOpdf

\else
\fi
\usepackage{mathtools}
\usepackage{amsmath}
\usepackage{amssymb}
\usepackage{cite}
\usepackage{graphicx}
\usepackage{multirow}
\usepackage{siunitx}
\usepackage{eufrak}
\usepackage{yfonts}
\DeclareMathOperator{\E}{\mathbb{E}}
\begin{document}
%
\title{{Statistical Distribution of Intensity Fluctuations for Underwater Wireless Optical Channels\\ in the Presence of Air Bubbles}}
\author{\IEEEauthorblockN{Mohammad Vahid Jamali, Pirazh Khorramshahi, Arvin Tashakori, Ata Chizari, Shadi Shahsavari,\\ Sajjad AbdollahRamezani, Masoome Fazelian, Sima Bahrani, and Jawad A. Salehi,\IEEEmembership{ Fellow,~IEEE}}
\IEEEauthorblockA{Optical Networks Research Lab, Department of Electrical Engineering, Sharif University of Technology, Tehran, Iran\\
Email: jasalehi@sharif.edu
}}



\maketitle
\begin{abstract}
 In this paper, we experimentally investigate the statistical distribution of intensity fluctuations for underwater wireless optical channels under different channel conditions, namely fresh and salty underwater channels with and without air bubbles. To do so, we first measure the received optical signal with a large number of samples.
Based on the normalized acquired data the channel coherence time and the fluctuations probability density function (PDF) are obtained for different channel scenarios.
Our experimental results show that salt attenuates the received signal while air bubbles mainly introduce severe intensity fluctuations. Moreover, we observe that log-normal distribution precisely fits the acquired data PDF for scintillation index ($\sigma^2_I$) values less than $0.1$, while Gamma-Gamma and K distributions aptly predict the intensity fluctuations for $\sigma^2_I>1$. Since neither of these distributions are capable of predicting the received irradiance for $0.1<\sigma^2_I<1$, we propose a combination of an exponential and a log-normal distributions to perfectly describe the acquired data PDF for such regimes of scintillation index.
\end{abstract}

\begin{keywords} 
Underwater wireless optical communications, intensity fluctuations, fading statistical distribution, channel coherence time.
\end{keywords}
\IEEEpeerreviewmaketitle
\section{Introduction}
The ever increasing demand for more secure, ecosystem friendly and broadband underwater communication and exploration necessitates comprehensive studies on underwater wireless optical communication (UWOC). Compared to its traditional counterpart, namely acoustic communication, UWOC provides better security, lower time delay and higher bandwidth \cite{tang2014impulse}, and is more compatible with underwater ecosystem. Despite all these, propagation of light under water is affected by three main degrading effects, namely absorption, scattering and turbulence, which limit the viable communication ranges of UWOC systems to typically less than $100$ \si{m}.

In recent years, many valuable researches have been carried out in the context of UWOC, from channel modeling to system design. In \cite{tang2014impulse,cox2012simulation,gabriel2013monte} Petzold's experimental data \cite{petzold1972volume,mobley1994light} are used to simulate the UWOC channel impulse response through Monte Carlo numerical method. However, these prior works only focused on the absorption and scattering effects of the channel, many relevant studies have been performed on the turbulence characterization. For example, in \cite{nikishov2000spectrum} an accurate power spectrum has been derived for fluctuations of turbulent seawater refractive index. Rytov method has been used in \cite{korotkova2012light} to evaluate the scintillation index of optical plane and spherical waves propagating in underwater turbulent medium. And in \cite{gerccekciouglu2014bit}, the on-axis scintillation index of a focused Gaussian beam is formulated for weak oceanic turbulence and log-normal distribution is considered for intensity fluctuations to evaluate the average bit error rate (BER) in such systems.

On the other hand, a cellular topology based on optical code division multiple access (OCDMA) technique has been recently proposed in \cite{akhoundi2015cellular} for underwater wireless optical networks, while the potential challenges and applications of such a network is discussed in \cite{akhoundi2016cellular}. Besides, beneficial applications of multi-hop transmission (serial relaying) on the up-and downlink performance of underwater users in an OCDMA network and also in point-to-point UWOC links have been investigated in \cite{jamali2015relay} and \cite{jamali2016multihop}, respectively, with respect to all the degrading effects of UWOC channels. Furthermore, the authors in \cite{jamali2015ber,jamali2015performanceMIMO} have focused on the turbulence-induced fading effects of UWOC channels and proposed multiple-input multiple-output (MIMO) transmission to mitigate turbulence effects through spatial diversity gain of MIMO technique.

Although, the statistical distribution of fading for free-space optical (FSO) channels is thoroughly investigated and characterized \cite{al2001mathematical,andrews2005laser,kashani2015novel}, its probability density function (PDF) for UWOC channels is not yet well determined.
Prior works on the context of UWOC mainly focused on weak oceanic turbulence and adopted the same statistical distribution as FSO channels, i.e., log-normal distribution for this regime of underwater turbulence \cite{yi2015underwater,ata2014scintillations,gerccekciouglu2014bit}. However, significant differences between the nature of atmospheric and oceanic optical channels may result in a different PDF for underwater fading statistics. The research in this paper is inspired by the need to more specifically study the intensity fluctuations through UWOC channels. In this paper, we experimentally investigate the effect of various phenomena on the fluctuations of the received optical signal within a laboratory water tank. Since the severity of optical turbulence rapidly increases with the link range, the short link range available with a laboratory water tank may typically result negligible fluctuations on the received optical signal. Therefore, at the first step, we artificially add air bubbles in our water tank to enlarge the channel's temporal fluctuations. Then, for each channel condition, we take sufficient samples from the received optical signal, normalize them to the received power mean, and plot the normalized samples histogram to obtain the intensity fluctuations' distribution and the channel coherence time; the time period in which the channel fading coefficient remains approximately constant.
It is worth mentioning that since many of underwater vehicles, divers and submarines generate air bubbles, study on the intensity fluctuations due to the presence of air bubbles has significant importance in precise UWOC channel modeling and system design.

The rest of the paper is organized as follows. In Section II the UWOC channel model is presented. The used evaluation metrics, and general statistical distributions for optical turbulence are briefly described in Sections III and IV, respectively. Section V provides the system model and experimental set up, Section VI presents experimental results for various scenarios and Section VII concludes the paper.
\section{UWOC Channel Description}
Propagation of light under water is affected by three major degrading effects, i.e., absorption, scattering and turbulence. In fact, photons of a propagating light wave may encounter with water molecules and particles. During this collision, energy of each photon may be lost thermally, which is specified as absorption process and is characterized by absorption coefficient $a(\lambda)$, where $\lambda$ is the wavelength of the propagating light wave. Furthermore, in the aforementioned collision process direction of each photon may be changed that is defined as scattering process and is determined by scattering coefficient $b(\lambda)$. Besides, total energy loss of non-scattered light is described by extinction coefficient $c(\lambda)=a(\lambda)+b(\lambda)$ \cite{mobley1994light}.

The channel fading-free impulse response (FFIR), $h_0(t)$, can be obtained through Monte Carlo simulations similar to \cite{tang2014impulse,gabriel2013monte,cox2012simulation}. Although, this FFIR includes both absorption and scattering effects, comprehensive characterization of the UWOC channel impulse response requires fading consideration as well. To do so, the channel FFIR can be multiplied by a fading coefficient \cite{jamali2015ber,andrews2005laser}. However, log-normal distribution is mainly used for fading coefficients' PDF \cite{yi2015underwater,gerccekciouglu2014bit,jamali2015ber,jamali2015relay}, inspired by the behaviour of optical turbulence in atmosphere, an accurate characterization of fading coefficients' statistical distribution necessitates more specific studies, that is carried out in this paper.

In order to specify the fading strength, it is common in the literature to define the scintillation index of a propagating light wave as \cite{andrews2005laser,korotkova2012light};
 \begin{align} \label{S.I.}
{\sigma }^2_I\left(r,d_0,\lambda \right)=\frac{\E\left[I^2(r,d_0,\lambda )\right] -{{\E}^2\left[I\left(r,d_0,\lambda \right)\right]}}{{{\E}^2\left[I\left(r,d_0,\lambda \right)\right]}},
\end{align}
in which $I(r,d_0,\lambda )$ is the instantaneous intensity at a point with position vector $\left(r,d_0\right)=(x,y,d_0)$, where $d_0$ is the propagation distance and $\E\left[I\right]$ denotes the expected value of the random variable (RV) $I$.
 \section{Evaluation Metrics}
 In order to investigate the UWOC channel's fading statistic, we first, experimentally measure the channel coherence time for different channel conditions. To do so, we measure the time in which the temporal covariance coefficient of irradiance, defined as Eq. \eqref{NTC}, remains approximately constant \cite{tang2013temporal,andrews2005laser}.
 \begin{align}\label{NTC}
b_{\tau,I}(d_0,\tau)=\frac{B_{\tau,I}(d_0,\tau)}{B_{\tau,I}(d_0,0)},
 \end{align}
 where $B_{\tau,I}(d_0,\tau)\stackrel{\triangle}{=}B_{I}(\boldsymbol{r_1},d_0,t_1;\boldsymbol{r_2},d_0,t_2)$ with $\boldsymbol{r_1}=\boldsymbol{r_2}$. Moreover, $\tau=t_1-t_2$, and $B_{I}(\boldsymbol{r_1},d_0,t_1;\boldsymbol{r_2},d_0,t_2)$ is defined as the covariance of irradiance for two points $\boldsymbol{r_1}$ and $\boldsymbol{r_2}$ at different time instants $t_1$ and $t_2$.
 
 As the next step of our work, we specify the channel fading statistical distribution. We use the following two well-known metrics to evaluate the accordance of different statistical distributions with the experimental data.
 \subsubsection{Root Mean Square Error (RMSE)}
 This metric determines how well the considered distribution predicts the experimental data, and is defined as;
 \begin{align}\label{RMSE}
 RMSE=\sqrt{\frac{1}{N}\sum_{i=1}^{N}\left(I_{m,i}-I_{p,i}\right)^2},
 \end{align}
 where $N$ is the number of measured samples, $I_{m,i}$ is the measured intensity through the experiment and $I_{p,i}$ is the predicted intensity by the considered distribution.
 \subsubsection{Goodness of Fit}
 This metric, which is also known as $R^2$ measure, is used to test the distribution's fitness and is expressed as;
 \begin{align}\label{R^2}
 R^2=1-\frac{SS_{reg}}{SS_{tot}},
 \end{align}
 in which $SS_{reg}$ is the considered distribution's sum of square errors, i.e., $SS_{reg}=\sum_{i=1}^{M}\left(f_{m,i}-f_{p,i}\right)^2$, where $M$ is the number of bins of the acquired data histogram, and $f_{m,i}$ and $f_{p,i}$ are, respectively, the measured and the predicted probability values for a given received intensity level. And $SS_{tot}$ is the sum of the squares of distances between the measured points and the mean of them, i.e., $SS_{tot}=\sum_{i=1}^{M}\left(f_{m,i}-\bar{f}\right)^2$, where $\bar{f}=\sum_{i=1}^{M}f_{m,i}/M$. As the value of the $R^2$ measure, for a given distribution, approaches its maximum (i.e., $1$), the distribution is considered to better fit the measured data.
 
 \section{General Statistical Distributions Used for Optical Turbulence Characterization}
  For the sake of brevity, we only take into account the following statistical distributions which are widely accepted for various regimes of atmospheric turbulence \cite{andrews2005laser}. Moreover, we propose a combined exponential and log-normal distribution to better characterize the fading statistics for $0.1<\sigma^2_I<1$.
 
 \textit{1) Log-Normal Distribution:} Such a distribution is commonly used in literature for weak atmospheric turbulence regime, characterized by $\sigma^2_I<1$. Let $\tilde{h}={\rm exp}(2X)$ be the channel fading coefficient ($\tilde{h}>0$) with log-normal PDF as \cite{andrews2005laser};
  \begin{equation} \label{log-normal}
 f_{\tilde{h}}({\tilde{h}})=\frac{1}{2\tilde{h} \sqrt{2\pi {\sigma }^2_X}}{\rm exp}\left(-\frac{{\left({{\rm ln}(\tilde{h})\ }-2{\mu }_X\right)}^2}{8{\sigma }^2_X}\right).
  \end{equation}
 Therefore, the fading log-amplitude $X=1/2\ln(\tilde{h})$ has a Gaussian distribution with mean ${\mu }_X$ and variance ${\sigma }^2_X$.
   To ensure that fading neither amplifies nor attenuates the average power, we normalize fading coefficients such that $\E[{\tilde{h}}]=1$, which implies  that ${{\mu }_X=-\sigma }^2_X$ \cite{andrews2005laser}. It can be shown that for log-normal fading the log-amplitude variance relates to the scintillation index as ${\sigma }^2_X=0.25\ln(1+\sigma^2_I)$ \cite{jamali2015performanceMIMO}.
 
  \textit{2) K Distribution:} This distribution is mainly used for strong atmospheric turbulence in which $\sigma^2_I\geq 1$. The PDF of K distribution is given by \cite{andrews2001laser};
  \begin{align}\label{K distribution}
  f_{\tilde{h}}(\tilde{h})=\frac{2\alpha}{\Gamma(\alpha)}\left(\alpha\tilde{h}\right)^{(\alpha-1)/2}K_{\alpha-1}\left(2\sqrt{\alpha\tilde{h}}~\!\right),
  \end{align}
  where $\Gamma(\alpha)$ is the Gamma function, $\alpha$ is a positive parameter which relates to the scintillation index as $\alpha=2/(\sigma^2_I-1)$ \cite{andrews2001laser}. And $K_p(x)$ is the $p$th order modified Bessel function of the second kind.
 
 \textit{3) Gamma-Gamma Distribution:} Such a statistical model, which factorizes the irradiance as the product of two independent random
 processes each with a Gamma PDF, has been introduced in literature to describe both small-scale and large-scale atmospheric
fluctuations. The PDF of Gamma-Gamma distribution is expressed as \cite{andrews2001laser};
\begin{align}\label{GG distribution}
  f_{\tilde{h}}(\tilde{h})=\frac{2(\alpha\beta)^{(\alpha+\beta)/2}}{\Gamma(\alpha)\Gamma(\beta)}(\tilde{h})^{\frac{(\alpha+\beta)}{2}-1}K_{\alpha-\beta}\left(2\sqrt{\alpha\beta\tilde{h}}~\!\right),
\end{align}
 in which $\alpha$ and $\beta$ are parameters related to effective atmospheric conditions. The scintillation index for Gamma-Gamma distribution is given by $\sigma^2_I=1/\alpha+1/\beta+1/\alpha\beta$ \cite{andrews2001laser}.

\textit{4) Proposed Combined Exponential and Log-Normal Distribution:}
When the received optical signal has a large dynamic range or equivalently the received signal lies either in large or small levels, general single-lobe distributions cannot specify the fading statistical distribution. In such circumstances, a two-lobe statistical distribution is required to interpret the channel intensity fluctuations. Therefore, we propose the combination of an exponential and a log-normal distributions with PDF of;
\begin{align}\label{exp+lognrm}
f_{\tilde{h}}({\tilde{h}})\!=\!\frac{k}{\gamma}\exp(-\tilde{h}/\gamma)\!+\!\frac{(1\!-\!k){\exp}\left(-{{\left({{\ln}(\tilde{h})\ }\!-{\mu }\right)}^2}/{2{\sigma }^2}\right)}{\tilde{h}\sqrt{2\pi {\sigma }^2}},
\end{align}
where, $k$ determines the proportion between the exponential and the log-normal distributions, $\gamma$ is the exponential distribution mean, and $\mu$ and $\sigma^2$ are the constants of the log-normal distribution. Fading coefficients normalization implies that $\E[\tilde{h}]=k\gamma+(1-k)\exp\left(\mu+{\sigma^2}/{2}\right)=1$. Moreover, the scintillation index and the $n$th ($n$ is a positive integer) moment for the above distribution can be obtained as $\sigma^2_I=2k\gamma^2+(1-k)\exp\left(2\mu+2{\sigma^2}\right)-1$ and $\E[{\tilde{h}}^n]=k\gamma^nn!+(1-k)\exp\left(n\mu+n^2{\sigma^2}/{2}\right)$, respectively.

\section{Experimental Setup Description}
\begin{figure}
\centering
\includegraphics[width=3.4in]{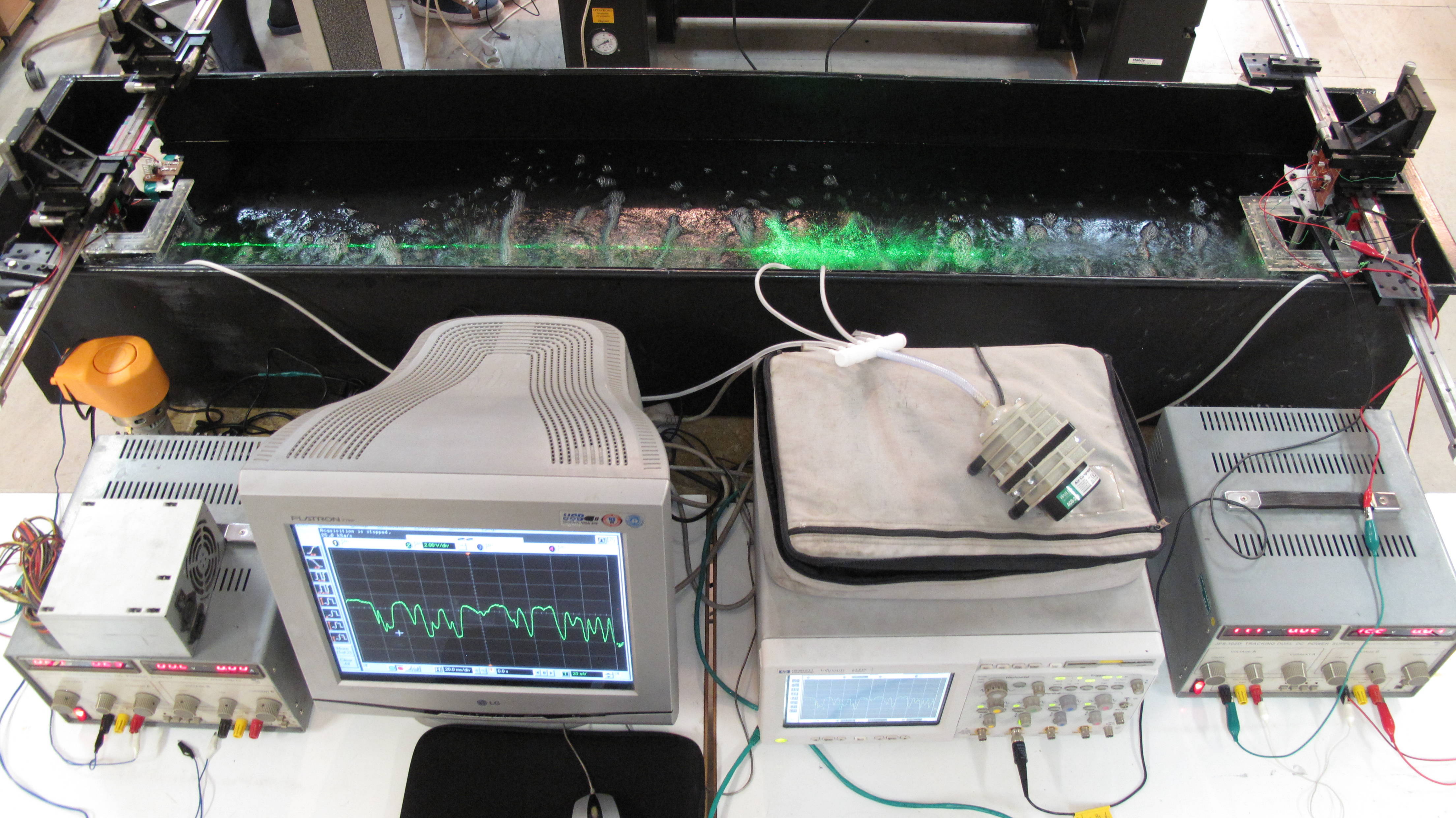}
\caption{Experimental setup.}
\label{fig:set_up}
\end{figure}
Our setup is implemented in a $30\times 40\times 200$ \si{cm^3} black colored water tank. In the transmitter side, a $532$ {\si nm} green laser diode with the maximum output power of $100$ {\si mW} is driven by a MOS transistor to ensure a constant optical irradiance. In the receiver side, an ultraviolate-visible photodetector (PD) is used; the PD is covered by a convex outer lens in order to collect the maximum amount of irradiance. Additionally, both laser and PD are
sealed by placing in transparent boxes. PD's generated current is amplified through AD840 operational amplifier. The magnified signal, which is proportional to the received optical power\footnote{Note that the PD's generated current $i(t)$ relates to the received optical power $p(t)$ as $i(t)=\eta p(t)/hf$, where $\eta$ is the PD's quantum efficiency, $h$ is the Planck's constant and $f$ is the light frequency. Moreover, the amplified signal is proportional to the PD's current and hence to the received optical signal. Therefore, normalized samples of the amplified photo-detected signal are good representatives of the received optical power normalized samples.}, is sampled and monitored by an HP Infiniium oscilloscope. For each test, we have collected $32768$ samples with the sampling rate of ${25}$ \si{{kSa}/{s}}. Along with our setup, an alcohol thermometer is attached to the lower side of the tank to ensure that the water temperature remains constant around $20$ degrees Celsius. Finally, to produce air bubbles, a tunable air blower with the maximum blowing capacity of $28$ \si{Litre/m} is employed.
Fig. \ref{fig:set_up} shows our experimental setup.
\section{Experimental Results}
\begin{figure*}
\centering
\includegraphics[width=7in]{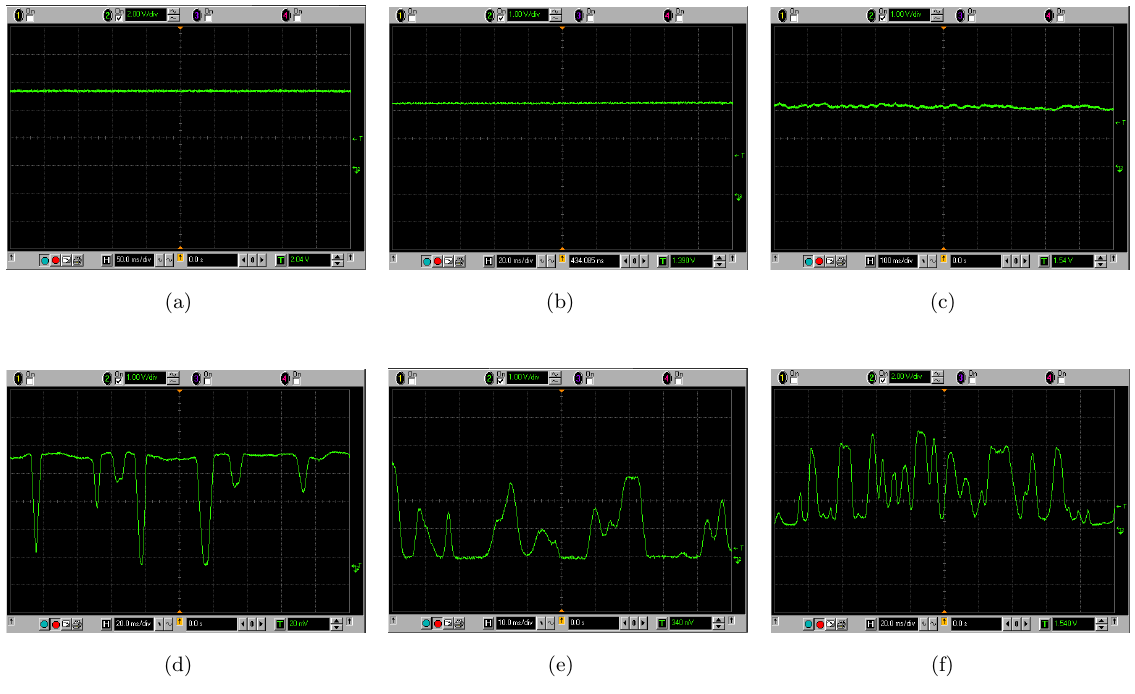}
\caption{Received optical signal through (a) free-space link with $\sigma^2_I=3.8181\times10^{-5}$; (b) fresh water link with $\sigma^2_I=6.1363\times10^{-5}$; (c) salty water link with $\sigma^2_I=9.8246\times10^{-4}$; (d) bubbly fresh water link with $\sigma^2_I=0.1015$; (e) bubbly fresh water link with $\sigma^2_I=2.1191$; (f) bubbly salty water link with $\sigma^2_I=1.0420$.}
\label{first_fig}
\end{figure*}
\begin{table*}
		\centering
		\caption{$R^2$ test and RMSE measure results for different statistical distributions and for various channel conditions, namely free-space (FS), fresh water (FW), salty water (SW), bubbly fresh water (BFW) with different values of $\sigma^2_I$, and bubbly salty water (BSW).}
		\begin{tabular}{||p{0.32in}||p{0.32in}||p{0.38in}|p{0.28in}||p{0.28in}|p{0.28in}||p{0.38in}|p{0.28in}|p{0.58in}||p{0.28in}|p{0.28in}|p{1.2in}||}
			\hline\hline
			\! Channel & \! $\sigma^2_I$& \multicolumn{2}{|c||}{Log-normal}&
			\multicolumn{2}{|c||}{K distribution}& \multicolumn{3}{|c||}{Gamma-Gamma distribution}&
			\multicolumn{3}{|c||}{Exponential+log-normal}\\
			\cline{3-12}
			&& ${R^2}$ & ${\rm RMSE}$ & ${R^2}$ & ${\rm RMSE}$ & ${R^2}$ & ${\rm RMSE}$ & Selected $(\alpha,\beta)$ set & ${R^2}$ & ${\rm RMSE}$ & Selected $(k,\gamma,\mu,\sigma^2)$ set \\ \hline \hline
			
			FS &$5.013 \times10^{-5}$ & $0.9956$ & $0.0100$ & $---$ & $---$ & $---$ & $---$ & $---$ & $0.9956$ & $0.0100$ & $(0,1,-2.5066\times10^{-5},5.0132\times10^{-5})$\\ \hline
			FW &$7.175 \times10^{-5}$ & $0.9672$ & $0.0120$ & $---$ & $---$ & $---$ & $---$ & $---$ & $0.9672$ & $0.0120$& $(0,1,-3.5876\times10^{-5},7.1752\times10^{-5})$\\ \hline
			SW &$9.824 \times10^{-4}$ & $0.9447$ & $0.0446$ & $---$ & $---$ & $---$ & $---$ & $---$ & $0.9447$ & $0.0445$ & $(0,1,-4.9098\times10^{-4},9.8196\times10^{-4})$\\ \hline
			BFW &$0.0011$ & $0.9582$ & $0.0462$ & $---$ & $---$ & $---$ & $---$ & $---$&$0.9582$&$0.0460$&$(0,1,-5.3513\times10^{-4},0.0011)$\\ \hline
			BFW &$0.0496$ & $0.4842$ & $0.3151$ & $---$ & $---$ & $0.4840$ & $0.3115$ & $(40,41.6224)$&$0.4842$&$0.3111$&$(0,1,-0.0242,0.0484)$\\ \hline
			BFW &$0.1015$ & $-0.0335$ & $0.4514$ & $---$ & $---$ & $0.0144$ & $0.3992$ & $(40,13.3983)$&$0.8180$&$0.5439$&$(0.3,0.5,0.185,0.005)$\\ \hline			
			BFW &$0.1477$ & $-0.1579$ & $0.5454$ & $---$ & $---$ & $-0.1056$ & $0.4816$ & $(40,8.3570)$ & $0.8085$ & $0.6290$& $(0.3, 0.35, 0.228, 0.0035)$ \\ \hline
			BFW &$0.2453$ & $-0.7065$ & $0.6985$ & $---$ & $---$ & $-0.5361$ & $0.6224$ & $(40,4.6532)$&$0.7385$&$0.7980$&$(0.58,0.6,0.35,0.003)$\\ \hline
						BFW &$0.5239$ & $-1.1541$ & $1.0221$ & $---$ & $---$ & $-0.7269$ & $0.9263$ & $(80,1.9800)$ & $0.4913$ &$1.0257$& $(0.301, 0.16, 0.63, 0.08)$\\ \hline
			BFW &$0.6456$ & $-0.9951$ & $1.1264$ & $---$ & $---$ & $-0.4050$ & $0.8035$ & $(60,1.6165)$ &$0.4463$&$1.1671$&$(0.58,0.38,0.5,0.15)$\\ \hline
			BFW &$0.7885$ & $-0.5169$ & $1.2429$ & $---$ & $---$ & $0.1673$ & $0.8880$ & $(60,1.3173)$&$0.7050$&$1.2552$&$(0.59,0.37,0.55,0.14)$\\ \hline
			BFW &$0.8173$ & $-0.2957$ & $1.2812$ & $---$ & $---$ & $0.6202$ & $0.9040$ & $(60,1.2699)$&$0.9026$&$1.3063$&$(0.71,0.48,0.64,0.18)$\\ \hline
			BFW &$0.9441$ & $-0.3048$ & $1.3747$ & $---$ & $---$ & $0.1141$ & $0.9717$ & $(60,1.0962)$&$0.7016$&$1.3421$&$(0.42,0.1,0.4,0.2)$\\ \hline
						BFW &$1.0652$ & $0.0045$ & $1.4510$ & $0.7126$ & $1.3937$ & $0.7324$ & $1.0321$ & $(60,0.9696)$& $0.9044$& $1.4788$& $(0.584, 0.32, 0.45, 0.33)$\\ \hline
						BFW &$1.1530$ & $ 0.2906$ & $1.5021$ & $0.8698$ & $1.4559$ & $0.9092$ & $1.0738$ & $(60,0.8947)$& $0.9506
						$ & $1.5154$ & $(0.571, 0.28, 0.47, 0.33)$\\ \hline
						BFW &$1.2838$ & $0.1944$ & $1.6076$ & $0.7990$ & $1.5141$ & $0.8918$ & $1.1330$ & $(60,0.8024)$& $0.9453$ & $1.5605$ & $(0.568, 0.29, 0.4824, 0.33)$\\ \hline
						BFW &$1.3448$ & $0.2890$ & $1.6295$ & $0.8430$ & $1.5672$ & $0.9136$ & $1.1596$ & $(60,0.7655)$&$0.9585$&$1.6589$&$(0.7,0.35,0.7,0.3)$\\ \hline
						BFW &$1.4312$ & $0.0865$ & $1.7249$ & $0.6812$ & $1.5786$ & $0.8325$ & $1.3298$ & $(60,0.7187)$ & $0.8470$ & $1.6212$ & $(0.568, 0.29, 0.4824, 0.36)$\\ \hline
						BFW &$3.5695$ & $0.1930$ & $2.5982$ & $0.6801$ & $5.3789$ & $0.7772$ & $1.9802$ & $(60,0.2862)$ & $0.6970$ & $2.4727$ & $(0.43, 0.1, 0, 1.01)$\\ \hline
			BFW &$12.0847$& $0.3801$& $3.4763$& $0.6090$& $3.4763$ & $0.6353$ &$3.4763$&$(0.4,0.3652)$&$0.9746$&$0.9746$&$(0.747,0.01,-0.02,1.76)$\\
			\hline
			BSW &$2.1776$& $0.2161$& $2.0896$& $0.7116$& $1.9565$ & $0.8687$ &$1.5831$&$(60,0.4705)$&$0.8325$&$1.9841$&$(0.5,0.14,0.25,0.65)$\\
						\hline
		\end{tabular}
\end{table*}

\begin{figure*}
\centering
\includegraphics[width=7in]{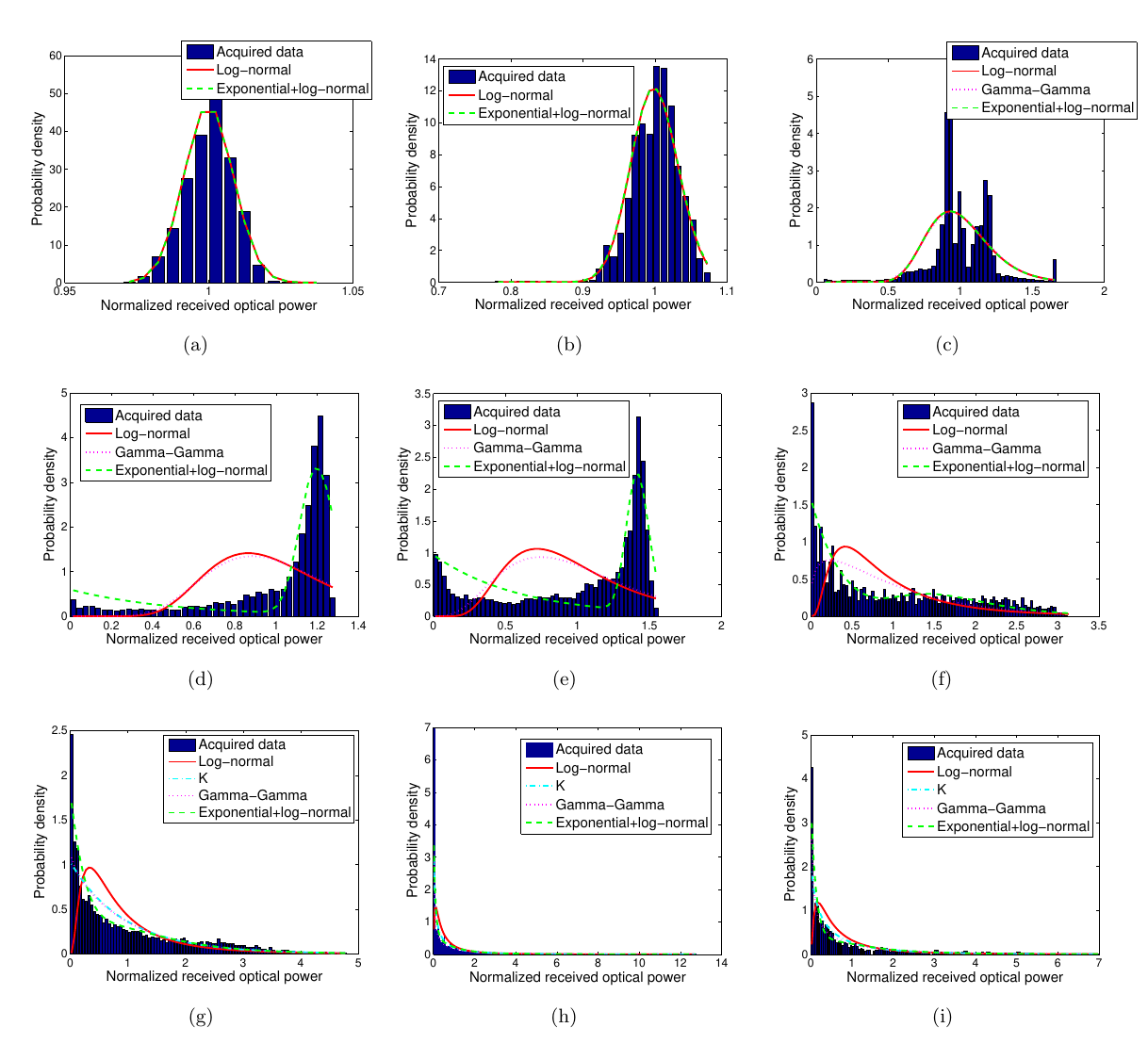}
\caption{Acquired data histogram along with different distributions' PDF for various channel scenarios, namely (a) fresh water link with $\sigma^2_I=7.175\times10^{-5}$; (b) bubbly fresh water link with $\sigma^2_I=0.0011$; (c) $0.0496$; (d) $0.1015$; (e) $0.2453$; (f) $0.7885$; (g) $1.0652$; (h) $3.5695$; and (i) bubbly salty water link with $\sigma^2_I=2.1776$.}
\label{hists}
\end{figure*}
In this section, we present our experimental results for the channel coherence time and the fading statistical distribution, under various underwater channel conditions. For each scenario we evaluate the validity of different distributions using the defined metrics in Section III. The considered channel conditions include free-space link as well as fresh and salty underwater channels with and without bubbles. For each scenario, we adjusted the transmitter laser power to ensure that a considerable power reaches the receiver.

Figs. \ref{first_fig}(a) and \ref{first_fig}(b) illustrate the received signal through the free-space and fresh water links, respectively. As it can be seen, for FSO and UWOC channels with low link ranges (like our water tank) the received signal is approximately constant over a large period of time, i.e., fading has a negligible effect on the performance of low-range FSO and UWOC systems. As a consequence, the channel impulse response for such scenarios can be thoroughly described by a deterministic FFIR.

In order to investigate salinity effects on the underwater optical channels, we added $700$ \si{g} salt to our water tank. We observed significant loss on the received signal in comparison to fresh water link. Therefore, we increased the transmitter power to observe a notable signal at the receiver, and hence to better investigate the fading fluctuations due to water salinity. Fig. \ref{first_fig}(c) shows the received signal through the salty water. As it is obvious, water salinity does not impose severe fluctuations on the propagating signal. We also used a water pump to flow water within the tank. Our extra experiments indicated that as the velocity of moving medium increases, the received signal experiences faster and more severe fluctuations.
This is mainly due to the fact that as the water flow increases, particles within the water move more rapidly and this causes more randomness on the transmitting signal path.

According to Figs. \ref{first_fig}(a)-(c), due to the low link range available by the laboratory water tank, the received signal fluctuations are typically negligible. In other words, optical turbulence manifests its effect at longer link ranges, i.e., for UWOC and FSO channels with typically $d_0>10$ \si{m} and $d_0>1$ \si{km} link ranges, respectively \cite{korotkova2012light,andrews2001laser}. Therefore, an air compressor is used to produce bubbles and hence to induce a severe fading on the channel. Figs. \ref{first_fig}(d) and \ref{first_fig}(e) illustrates the received signal through the bubbly fresh water, while Fig. \ref{first_fig}(f) shows the same result for the bubbly and salty underwater link. As it is obvious from these figures, the presence of bubbles within the channel causes the received signal to severely fluctuate. This is mainly due to the random arrangement of bubbles encountering the propagating signal across the channel, that causes propagating photons to randomly scatter in different directions and leave their direct path.

Table I summarizes the results of $R^2$ test and RMSE measure for different statistical distributions and for various channel conditions\footnote{It is worth mentioning that since for K distribution $\sigma^2_I=1+2/\alpha$ and $\alpha$ is a positive quantity, the evaluation of K distribution is only possible when $\sigma^2_I>1$. Moreover, for very small values of $\sigma^2_I$ the the Gamma-Gamma distribution parameters, i.e., $\alpha$ and $\beta$ have large values hampering the analytical evaluation of Gamma-Gamma distribution for such regimes of scintillation index.}. As it can be seen, for very small values of scintillation index, i.e., $\sigma^2_I<0.01$ log-normal distribution has an excellent goodness of fit and a negligible RMSE. However, both log-normal and Gamma-Gamma distributions acceptably predict the irradinace fluctuations when $\sigma^2_I<0.1$, as $\sigma^2_I$ increases and approaches $0.1$ both of these distributions lose their accuracy. Meanwhile, both K and Gamma-Gamma distributions can aptly fit the experimental data for strong turbulent channels, characterized by $\sigma^2_I>1$, while log-normal distribution fails to describe the intensity fluctuations for such a regime of scintillation index. Fig. \ref{hists} better illustrates the fitness of different statistical distributions with the acquired data histograms for various channel scenarios.
Based on the summarized results of Table I, neither of the discussed general statistical distributions are capable of predicting the intensity fluctuations when $0.1<\sigma^2_I<1$. As the received intensity shapes in Figs. \ref{first_fig}(d)-(f) and the histograms of the acquired data in Figs. \ref{hists}(d)-(f) show the presence of air bubbles causes the received intensity to mainly lie either in large or small values. Hence, typical single-lobe distributions cannot appropriately fit the experimental data and generally a two-lobe statistical distribution is required. Therefore, we proposed a combined exponential and log-normal distribution as discussed in Section IV. As Table I and Fig. \ref{hists} illustrate the proposed distribution excellently fits the acquired experimental data for all the regions of scintillation index with an acceptable RMSE.

\begin{figure}
\centering
\includegraphics[width=3.4in]{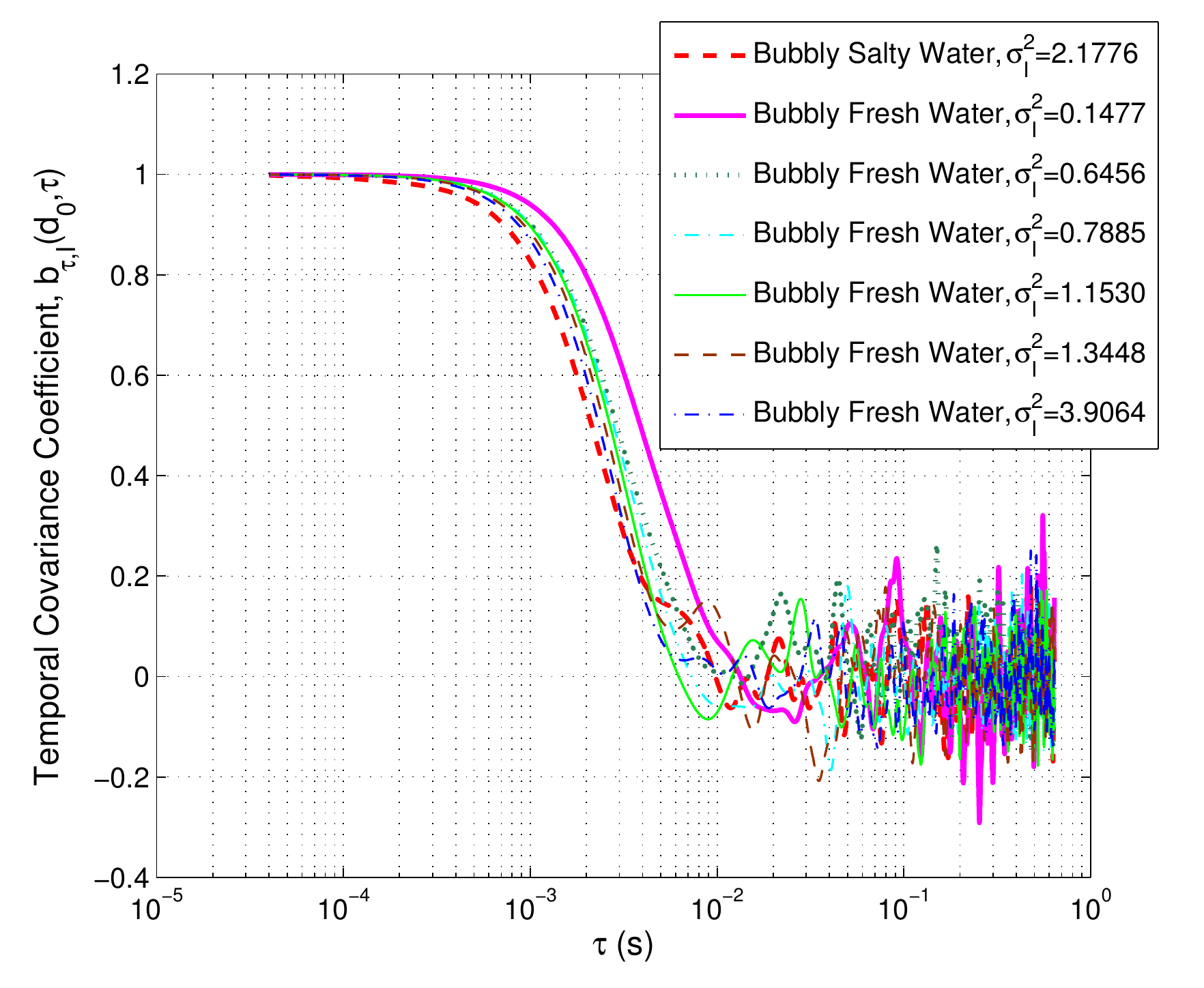}
\caption{Temporal covariance coefficient
for various channel scenarios.}
\label{TCC}
\end{figure}
Fig. \ref{TCC} illustrates the temporal covariance coefficient of irradiance for different scenarios. As it can be seen, even in strongest fading condition of bubbly channels the aforementioned coefficient is larger than $10^{-3}$ seconds confirming the flat fading, i.e., the same fading coefficient over thousands up to millions of consecutive bits. Moreover, this figure shows that the channel temporal variation slightly increases with the increase on the scintillation index value, i.e, the higher the value of $\sigma^2_I$, the less the value of $b_{\tau,I}(d_0,\tau)$.
\section{Conclusions}
In this paper, we experimentally investigated the statistical distribution of intensity fluctuations for underwater wireless optical channels under different channel conditions. Our experimental results showed that salt attenuates the received signal while air bubbles mainly introduce severe intensity fluctuations. Moreover, we observed that log-normal distribution precisely fits the acquired data PDF for scintillation index values less than $0.1$, while Gamma-Gamma and K distributions aptly predict the intensity fluctuations for $\sigma^2_I>1$. Meanwhile, neither of these distributions are capable of predicting the received irradiance for $0.1<\sigma^2_I<1$. Therefore, we proposed a combination of an exponential and a log-normal distributions to perfectly describe the acquired data PDF for such regimes of scintillation index. Furthermore, our study on the channel coherence time have shown larger values than $10^{-3}$ seconds for the channel temporal covariance coefficient implying flat fading UWOC channels.


%


\end{document}